\documentstyle[11pt,newpasp,twoside,epsf]{article}
\def\kms{km s$^{-1}$}

\def\degree{^\circ}
\def\eg{e.g.,\ }

\def\edcomment#1{\iffalse\marginpar{\raggedright\sl#1\/}\else\relax\fi}
\reversemarginpar

\begin{document}
\title{Kinematics of Planetary Nebulae in M51's Tidal Tail}
\author{John J. Feldmeier, J. Christopher Mihos}
\affil{Case Western Reserve University, 10900 Euclid Ave. 
Cleveland, OH 44106, U.S.A.}
\author{Patrick R. Durrell, Robin Ciardullo}
\affil{Penn State University, 525 Davey Lab, University Park, PA 16802, U.S.A.}
\author{George H. Jacoby}
\affil{WIYN Observatory, P.O. Box 26732, Tucson AZ 85726, U.S.A.}

\begin{abstract}
The galaxy pair NGC 5194/95 (M51) is one of the closest and best known
interacting systems.  Despite its notoriety, however, many of its features
are not well studied.  Extending westward from NGC~5195 is a 
low surface brightness tidal tail, which can only be
seen in deep broadband exposures.  Our previous [O~III] $\lambda 5007$ 
planetary nebulae (PN) survey of M51 recovered this tidal tail, 
and presented us with a opportunity to study the kinematics of a 
galaxy interaction in progress.  We report the results of a spectroscopy 
survey of the 
PN, aimed at determining their kinematic properties.  We then use 
these data to constrain new self-consistent numerical models of the system.
\end{abstract}

M51 is one of the most famous interacting galaxy systems, but 
after years of modeling (\eg Toomre \& Toomre 1972; Hernquist 1990; 
Salo \& Laurikainen 2000), a full description 
of the encounter is still not well understood.  
To study this system in more detail, we completed 
an [O~III] survey for planetary nebulae (PN) in M51 (Feldmeier, 
Ciardullo, \& Jacoby 1997).  Surprisingly, 
we found a large number of PN to the west of the 
secondary galaxy NGC~5195 in a tidal-tail like structure 
(Fig.~1, left). Although this feature had been detected 
previously with deep photographic imaging (Burkhead 1978), 
it has remained unstudied until now.

To study the kinematics of the M51 system, we used the
Hydra fiber spectrograph on the WIYN 3.5-m telescope to measure the
radial velocities of 36 of the galaxies' halo PN.  By comparing 
the spectra of identical PN taken in multiple setups, we find 
that the velocity uncertainties are less than 17 \kms.  We find a 
complex velocity distribution in the tidal tail, with multiple
components that can be attributed to each galaxy. 

\begin{figure}
\plottwo{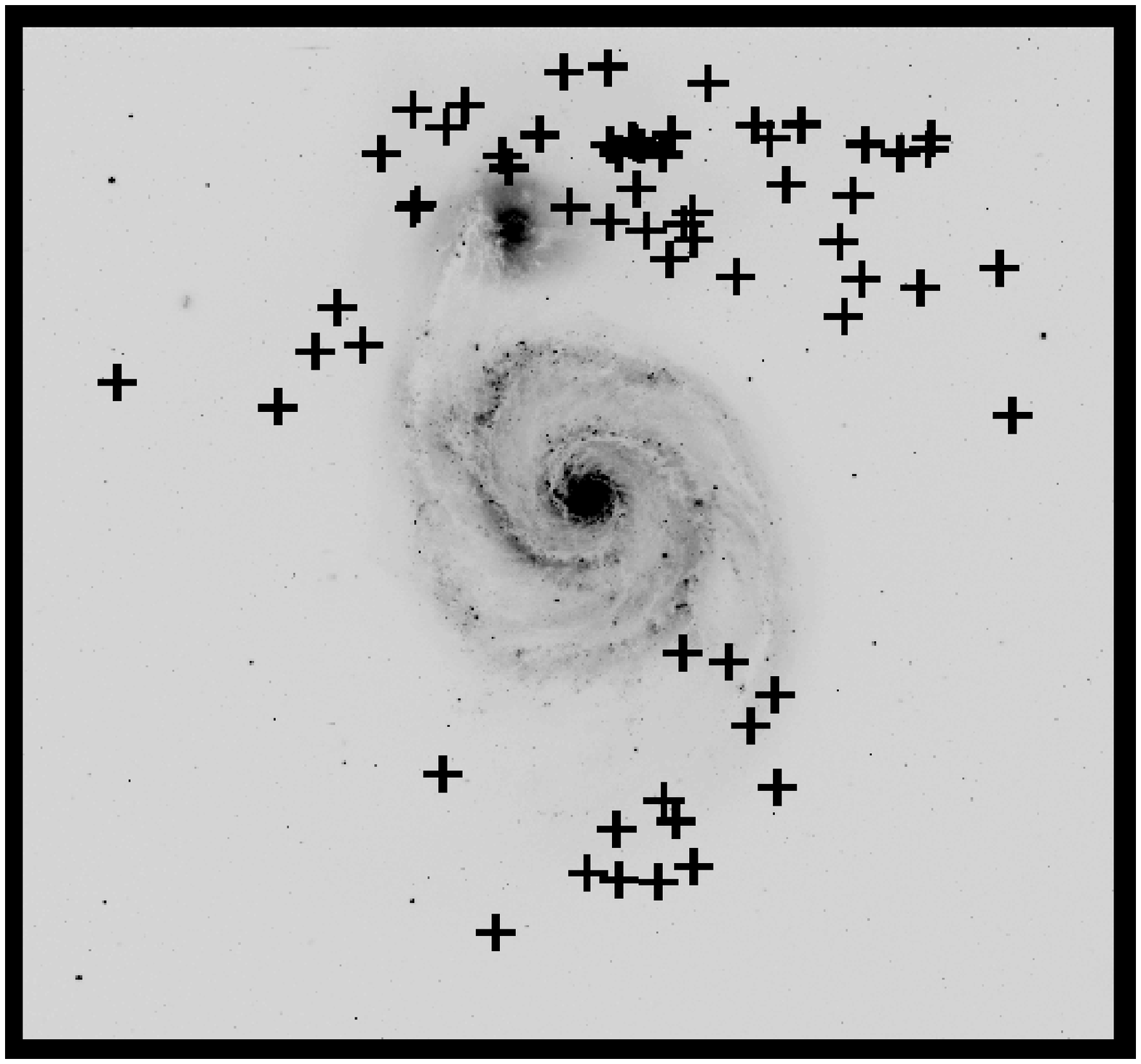}{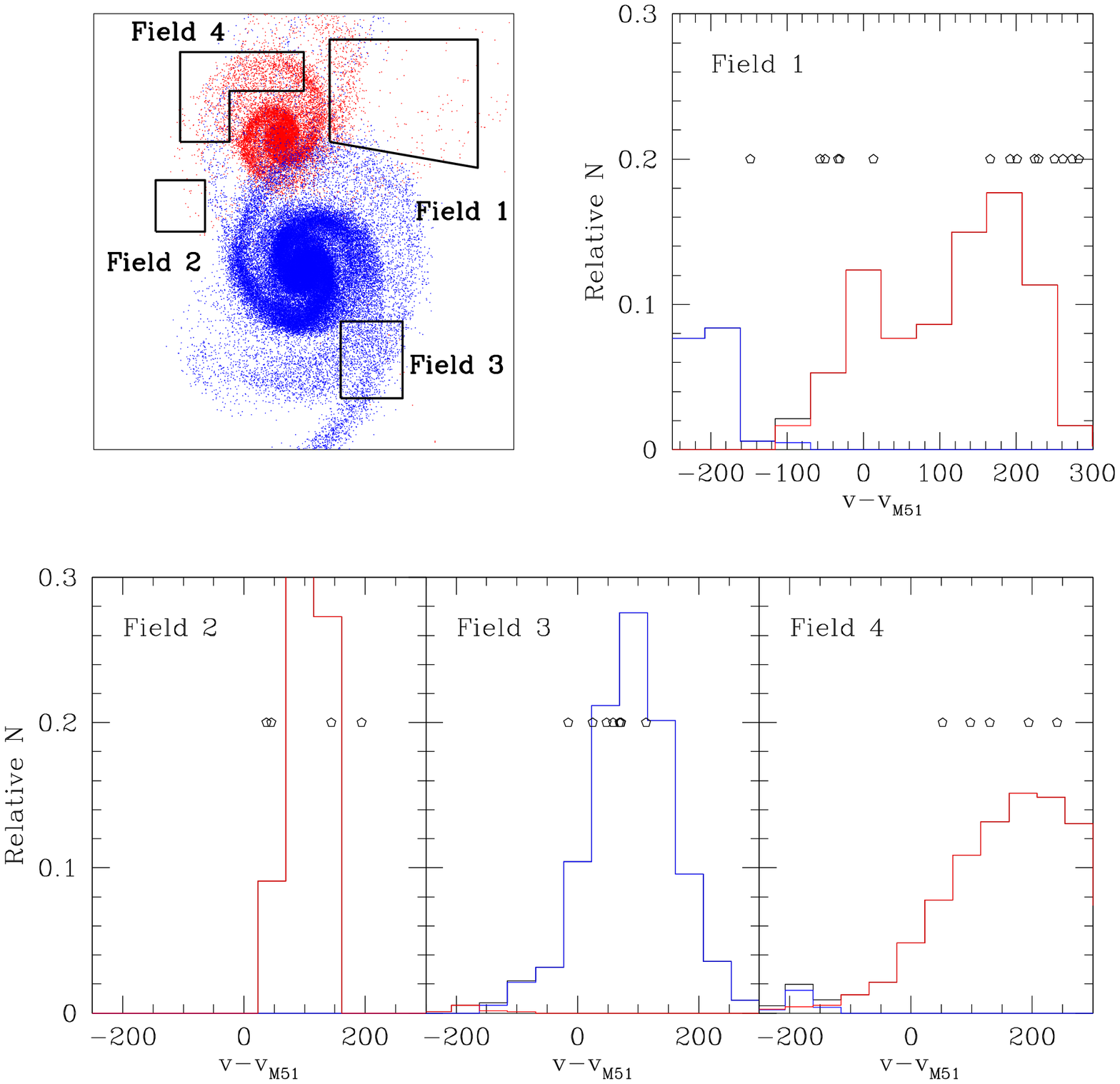}
\caption{On the left is our [O~III] $\lambda$ 5007 image of the M51 
system; the 64 PN are labeled by crosses.  The diagrams on the right 
compare the results of our N-body models (the histograms) with 
the observed PN velocities (open pentagons).}
\end{figure}

To understand the kinematic properties of the PN, 
we constructed an N-body model of the M51 system.  We began by using 
earlier simulations (\eg Hernquist 1990) which suggested 
that the encounter involves a mass ratio of 2:1, an elliptical 
orbit ($e \sim 0.7$), 
and a relatively high ($i=70\degree$) inclination of NGC 5195's 
disk to the orbital plane.  We then varied the orientation of the 
companion to produce the optical morphology and general PN kinematics. 
We added a bulge to each galaxy (B:D = 1/3) to suppress the bar 
instability in the disks. The model galaxies 
were constructed in the manner described by Hernquist (1993)
and evolved in a self-consistent manner using TREECODE (Hernquist 
1987).  We found that in the best-fitting model the companion 
is slightly retrograde ($i=110\degree$, $\omega=30\degree$) 
and the system is viewed approximately 300 Myr after the 
initial encounter.

To analyze the PN velocities, we divided the galaxy model 
into different fields and constructed radial velocity histograms 
of the disk particles within each region (Fig.~1, right).  
The tidal tail region (Field 1) 
shows a broad, complex, and multi-modal structure arising from an 
overlapping of PN in the disk of NGC 5195, and in the tidal tails of 
both M51 and NGC 5195.  Fields 2--4 are quieter kinematically, and 
sample other tidal features in the system.  With very little 
fine tuning, we can reconstruct the velocity distribution of the 
observed PN using a relatively simple interaction model.



\begin{references}

Burkhead, M.S. 1978, \apjs, 38, 147

Feldmeier, J., Ciardullo, R. \& Jacoby, G. 1997, \apj, 479, 231

Hernquist, L. 1987, \apjs, 64, 715

Hernquist, L. 1990, in Dynamics and interactions of galaxies, ed. R. 
Wielen (New York: Springer-Verlag), 108. 

Hernquist, L. 1993, \apjs, 86, 389 

Salo, H. \& Laurikainen, E. 2000, \mnras, 319, 39

Toomre, A. \& Toomre, J. 1972, \apj, 178, 623

\end{references}
\end{document}